\documentclass[twocolumn,twoside,slac_two]{revtex4}
\usepackage{graphicx}
\usepackage{fancyhdr}
\begin{document}

\title{New unidentified H.E.S.S. Galactic sources}

%

\author{O. Tibolla, R.C.G. Chaves, W. Domainko}
\affiliation{Max-Planck-Institut f\"ur Kernphysik, Heidelberg, P.O. Box 103980, D69029 Heidelberg, Germany}
\author{O. de Jager}
\affiliation{Unit for Space Physics, North-West University, Potchefstroom 2520, South Africa}
\author{S. Kaufmann, S. Wagner}
\affiliation{Landessternwarte, Universit\"at Heidelberg, K\"onigstuhl, D 69117 Heidelberg, Germany}
\author{N. Komin}
\affiliation{CEA, Irfu, SPP, Centre de Saclay, F-91191 Gif-sur-Yvette, France}
\author{K. Kosack}
\affiliation{CEA, Irfu, SAP, Centre de Saclay, F-91191 Gif-sur-Yvette, France}
\author{A. Fiasson}
\affiliation{Laboratoire d'Annecy-le-Vieux de Physique des Particules, CNRS/IN2P3, \\ 9 Chemin de Bellevue - BP 110 F-74941 Annecy-le-Vieux Cedex, France}
\author{M. Renaud}
\affiliation{AstroParticule et Cosmologie, CNRS-UMR 7164, Universit\`e Paris 7 Denis Diderot, F-75205 Paris, France}
\author{on behalf of the H.E.S.S. collaboration}
\affiliation{..\\}
\begin{abstract}
H.E.S.S. is one of the most sensitive instruments in the very high energy (VHE; $>$ 100 GeV) gamma-ray domain and has revealed many new sources along the Galactic Plane. After the successful first VHE Galactic Plane Survey of 2004, H.E.S.S. has continued and extended that survey in 2005--2008, discovering a number of new sources, many of which are unidentified. \\
Some of the unidentified H.E.S.S. sources have several positional counterparts and hence several different possible scenarios for the origin of the VHE gamma-ray emission; their identification remains unclear. Others have so far no counterparts at any other wavelength. Particularly, the lack of an X-ray counterpart puts serious constraints on emission models. \\
Several newly discovered and still unidentified VHE sources are reported here.
\end{abstract}
\maketitle

\thispagestyle{fancy}


\section{Introduction}
Very high energy (VHE, $> 10^{11}$ eV) particles can be traced within our Galaxy by a combination of non-thermal X-ray emission and VHE gamma-ray emission via leptonic (i.e. Inverse Compton scattering of electrons, Bremsstrahlung and synchrotron radiation) or hadronic (i.e. the decay of charged and neutral pions, due to interactions of energetic hadrons) processes.

The High Energy Stereoscopic System (H.E.S.S.) detects VHE $\gamma$-rays above an energy threshold of $\sim$100 GeV and up to $\sim$100 TeV with a typical energy resolution of 15\% per photon. The angular resolution is $\sim 0.1^{\circ}$ per event, allowing a positional error better than 40'' for a point source detected with a statistical significance 6 $\sigma$. The H.E.S.S. field of view is almost $5^{\circ}$ in diameter with a point source sensitivity of $<2.0 \times 10^{-13}$ ergs cm$^{-2}$ s$^{-1}$ ($\sim$1\% of the Crab Nebula) for a  5 $\sigma$ detection in 25 hours of observations \cite{1}.

\section{New unidentified H.E.S.S. sources}
After the successful first survey of 2004 \cite{2}, H.E.S.S. extended the survey in 2005-2008 \cite{ryan}, leading to the discovery of several new VHE gamma-ray sources. 
Of these, several have been associated with Supenovae Remnants (SNRs; such as CTB 37A, CTB 37B, RCW 86, Kes 75), some of them are candidate PWNe (such as HESS J1356-654 and HESS J1849-000), however the rest remain unidentified.

The understanding of the nature of the unidentified sources is a crucial point, since they represent a large fraction of the high energy and very high energy source population: almost 50\% of the H.E.S.S. Galactic sources are still unidentified; more than 60\% of EGRET sources were
unidentified \cite{3EG} and also almost 50\% of Fermi LAT sources are unidentified \cite{bsl}. \\
Unidentified sources could point to unknown and potentially new phenomena. And also in the case that they would represent particular unknown stages of known phenomena, their study will allow us to put serious constraints on the current astrophysical models and hence on their physical emission models. E.g. in the case that the VHE unidentified sources would be, as suggested by \cite{ancientPWN}, ancient PWNe, a new stage of evolution of PWNe is discovered and studied, putting serious constraints on the model of emission.

Here five recently discovered VHE gamma-ray sources, that are still unidentified, will be discussed, showing their morphology, their spectrum and providing evidence for possible counterparts: {\bf HESS J1507-622} (discussed separately in \cite{1507Fermi}), {\bf HESS J1848-018}, {\bf HESS J1503-582} and a particular emphasis is dedicated to the new sources in the Galactic Center region, which is very important for Fermi LAT studies of diffuse emission or dark matter ({\bf HESS J1741-302} and {\bf HESS J1745-303}).

\section{HESS J1503-582}
\begin{figure}[t]
   \centering
\includegraphics[width=\columnwidth]{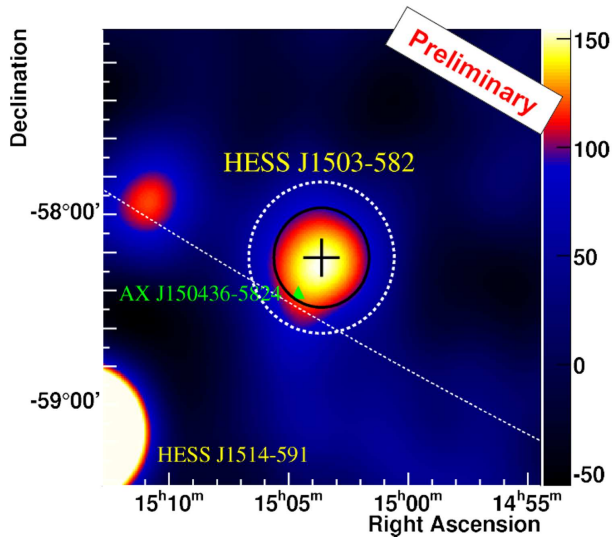}
  	\caption{Excess count map of HESS J1503-582. The white dashed line represents the Galactic Plane. Superimposed the X-ray source AX J150436-5824 tentatively classified as a Cataclysmic Variable. In the field of view there is also the pulsar PSR J1502-5828, but it has a very low spin-down luminosity ($3.3 \times 10^{31} (d / 12 \mathrm{kpc})^{-2} \mathrm{erg / s~kpc^2}$), and is not powerful enough to explain a PWN scenario.}
         \label{1503}
   \end{figure}

The VHE gamma-ray emitter HESS~J1503-582 (Figure \ref{1503}) has been recently discovered by H.E.S.S. \cite{RenaudGamma08}, it is still unidentified, i.e. it does not have any of the typical counterparts at lower energies (e.g. SNRs, energetic pulsars, or bright \emph{Fermi} gamma-ray sources).  However, it is now considered unique, since it is the first VHE gamma-ray source that appears to be associated with a \emph{forbidden-velocity wing} (FVW).  FVWs are faint HI 21 cm emission line structures which are visible at velocities which deviate from the canonical Galactic rotation curve, suggesting associated dynamical phenomena \cite{KangKoo2007}.  Deeper observations of HESS J1503-582 are scheduled in order to further investigate this new source of VHE gamma-rays.

HESS J1503-582, visible in Figure \ref{1503}, shows significance of $\sim$6$\sigma$ in $\sim$24 hours of effective exposure. The spectrum can be fit well by a power law with index $\Gamma$~=~2.4~$\pm$~0.4$_{\mathrm{stat}}$~$\pm$~0.2$_{\mathrm{syst}}$ with a flux above 1 TeV of $\sim 6 \times 10^{-12}$ cm$^{-2}$ s$^{-1}$ ($\sim$6 \% of the Crab flux).

\section{HESS J1848-018}
\begin{figure}[t]
   \centering
\includegraphics[width=\columnwidth]{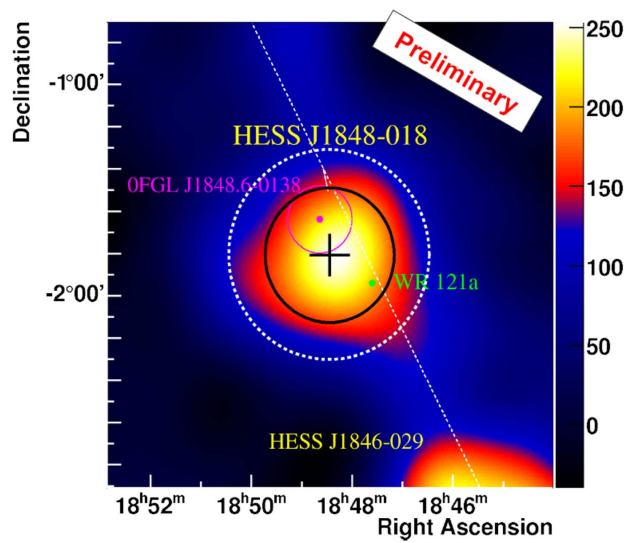}
 	\caption{Excess count map of HESS J1848-018. The white dashed line represents the Galactic Plane. The white circle denotes the source's intrinsic rms size of 0.32$^{\circ}$~$\pm$~0.02$^{\circ}$. The position of WR~121a is marked in green and in magenta is shown the position of the Fermi bright source OFGL J1848.6-0138 and its radius of 95\% confidence region \cite{bsl}.}
         \label{1848}
   \end{figure}

HESS~J1848-018 is an extended VHE gamma-ray source which was recently detected in the H.E.S.S. Galactic Plane Survey with a statistical significance of $\sim$9$\sigma$ in $\sim$50 hours of effective exposure \cite{Chaves08}.  Figure \ref{1848} shows the excess map of the recently discovered source HESS J1848-018 \cite{Chaves08}. Its differential energy spectrum is well fit by a power law with index $\Gamma$~=~2.8~$\pm$~0.2$_{\mathrm{stat}}$~$\pm$~0.2$_{\mathrm{syst}}$ and it has an integrated flux above 1 TeV of $\sim$2~$\times$~10$^{-12}$~erg~cm$^{-2}$~s$^{-1}$, corresponding to $\sim$8\% of the Crab nebula. 

HESS J1848-018 is located along the Scutum-Crux spiral arm tangent and is in the direction of, but slightly offset from, the star-forming region W~43, which hosts a giant HII region (G30.8$-$0.2), a giant molecular cloud, and the Wolf-Rayet (WR) star WR~121a in the central stellar cluster.  If HESS~J1848-018 is indeed associated with W~43, it would be only the second known case, after Westerlund 2 \cite{Wd2}, of VHE gamma-ray emission associated with a star-forming region and WR star. \\
The complex, multi-wavelength morphology of HESS J1848-018 is currently under investigation.

\section{Galactic Center region: HESS J1745-303 and HESS J1741-302}
The discovery of new VHE sources close to the Galactic Center is very important for the studies about diffuse Galactic emission and dark matter and will play a crucial role in the out-coming Fermi LAT studies.

\subsection{HESS J1745-303}
HESS J1745-303 is not a new source \cite{2}, but it has been recently observed in VHE gamma-rays in detail by H.E.S.S. and in X-rays by \emph{XMM-Newton} \cite{1745}. 

   \begin{figure}
   \centering
  \includegraphics[width=\columnwidth]{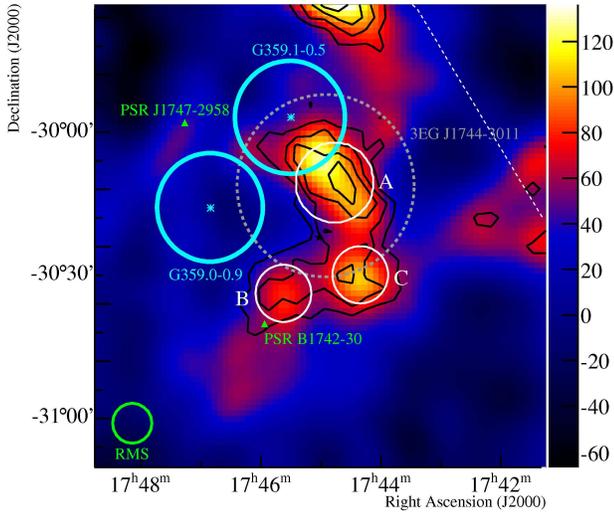}
  	\caption{Excess count map (smoothed with Gaussian of $0.07^{\circ}$ radius) and significant contours of HESS  J1745-303 \cite{1745}. The white dashed line represents the Galactic Plane. Superimposed there are the two SNRs G359.1-0.5 and G359.0-0.9, the two pulsars PSR B1742-30 and PSR J1747-2958, and the EGRET source 3EG 1744-3011.}
         \label{1745fig}
   \end{figure}
~

Figure \ref{1745fig} shows the image of gamma-ray excess counts of HESS J1745-303 smoothed with a radius of  $0.07^{\circ}$. Overlaid on the image are the contours of 4$\sigma$, 5$\sigma$, 6$\sigma$, 7$\sigma$ of significance.

HESS J1745-303 is detected with a significance of 10.2$\sigma$ in 79 hours of observation. The spectrum can be fit well by a power law with index $\Gamma = 2.7 \pm 0.1_{stat} \pm 0.2_{sys}$ and a flux above 1 TeV of $(1.63 \pm 0.16) \times 10^{-12}$ cm$^{-2}$ s$^{-1}$.  \\
The region labeled A in Figure \ref{1745fig} is thought to be associated with the interaction of the SNR G359.1-0.5 \cite{downes} with $^{12}$CO molecular clouds \cite{bitran} spatially coincident with the peak of the VHE gamma-ray emission. The part labeled B in Figure \ref{1745fig} is more likely to be a PWN powered by the pulsar PSR B1742-30 ($\dot{E}/D^2 = 2 \times 10^{33}$ erg s$^{-1}$ kpc$^{-2}$), which requires a conversion efficiency from rotational kinetic energy to gamma-rays of $\sim32 \%$ to produce the entire VHE emission.
HESS J1745-303 is also spatially coincident with the EGRET source 3EG 1744-3011 \cite{3EG}.

\subsection{HESS J1741-302}
The discovery of a faint ($\sim$1\% of the Crab flux) source, HESS J1741-302, has been recently announced \cite{1741}, one of the faintest newly discovered sources, at the lower end of the H.E.S.S. sensitivity.

   \begin{figure}
   \centering
\includegraphics[width=\columnwidth]{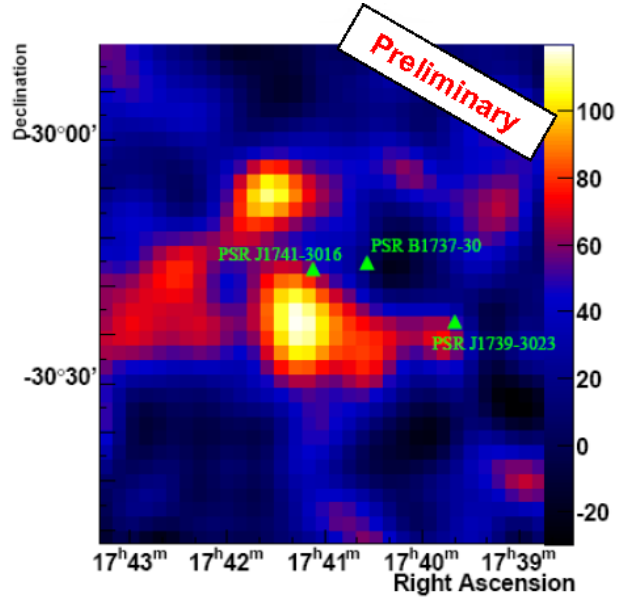}
  	\caption{Gamma-ray acceptance corrected skymap of HESS J1741-302 \cite{1741}, obtained with \emph{hard analysis cuts} and a Gaussian smoothing radius of $0.05^{\circ}$, with superimposed the positions of PSR B1737-30 \cite{6} and PSR J1741-3016 \cite{7}.}
         \label{1741fig}
   \end{figure}

\begin{figure*}[t]
  \includegraphics[width=120mm]{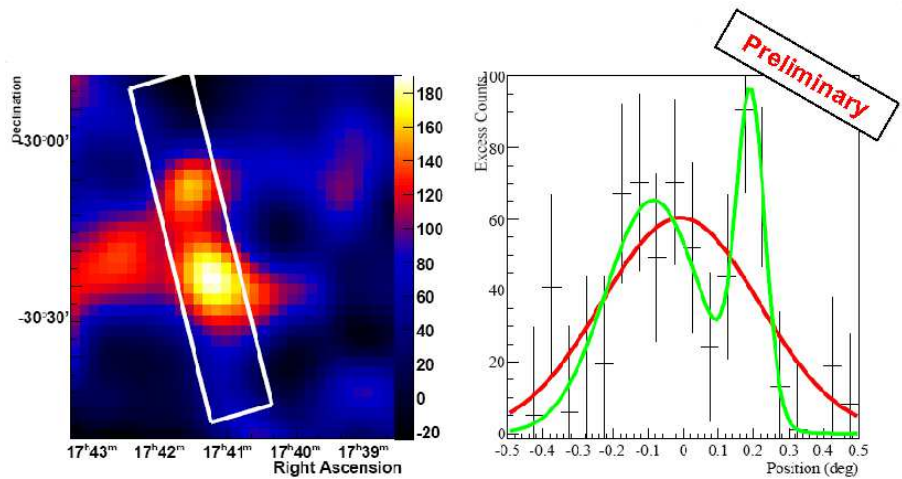}
  \caption{On the left: gamma-ray acceptance corrected skymaps of HESS J1741-302 with superimposed the chosen slice connecting the two hotspots. On the right: the excess counts in the slice are plotted and fitted with a Gaussian (in red, with ${\chi}^{2}/d.o.f.=30.18/17$) and with a double Gaussian distribution (in green, with ${\chi}^{2}/d.o.f.=10.24/14$) \cite{1741}.}
  \label{nostat}
\end{figure*}

The discovery peak significance is 8.1 $\sigma$ in 143.5 hours of observations. With \emph{standard analysis cuts} \cite{berge}, HESS J1741-302 appears as an irregular blob. With \emph{hard analysis cuts} (200 p.e.) \cite{berge} resulting in improved angular resolution, improved background rejection and higher energy threshold, two apparent hot spots start to appear in the map (Figure \ref{1741fig}); however Figure \ref{nostat} shows that current statistics do not allow detailed statements about source morphology, and that the hot spots are consistent with statistical fluctuations within the source. \\
The preliminary spectrum can be fit well by a power law with index $\Gamma = 2.78 \pm 0.24_{stat} \pm 0.20_{sys}$ with a flux above 1 TeV of $(6.3 \pm 1.3_{stat} \pm 1.1_{sys}) \times 10^{-13}$ cm$^{-2}$ s$^{-1}$.

In other sources in the galactic center region (e.g. HESS J1745-303 \cite{1745} and the GC diffuse emission \cite{GC}), there is evidence that interactions of cosmic rays with Molecular Clouds (MCs) play a role in VHE gamma-ray production. Such an association in this case is not clear, though in some velocity ranges there is some weak molecular emission coincident with the H.E.S.S. source \cite{4}.
Another suitable scenario could be an offset Pulsar Wind Nebula powered by the somewhat offset but relatively powerful pulsar PSR B1737-30 \cite{6} (spin-down luminosity $7.7 \times 10^{33} ~ \mathrm{\frac{erg}{s \cdot kpc^{2}}}$). This pulsar has a spin-down luminosity that is almost one order of magnitude fainter than other VHE gamma-ray emitting pulsars (e.g. like HESS J1825-137 \cite{pwn}), which have typical VHE fluxes of $\sim$10\% of the Crab. Given that the estimated flux of HESS J1741-303 is approximately 1\% of the Crab, a weak PWN scenario is plausible.
The brightest hot spot is spatially coincident with the pulsar PSR J1741-3016 \cite{7}, which is, however, probably too faint (spin-down luminosity $2.1 \times 10^{30} ~ \mathrm{\frac{erg}{s \cdot kpc^{2}}}$) to sustain a PWN scenario.

\section{Conclusions}
Five recently discovered H.E.S.S. unidentified sources and their possible counterparts have been discussed. All the sources are located in the Galactic plane (or slightly off the plane) and appear extended in VHE gamma rays. Since these sources have so far no clear counterpart in lower-energy wavebands, further multi-wavelength studies are required in order to understand the emission mechanism powering them.

\section{Acknowledgements}
The support of the Namibian authorities and of the University of Namibia in facilitating the construction and operation of HESS is gratefully acknowledged, as is the support by the German Ministry  for Education and Research (BMBF), the Max Planck Society, the French Ministry for Research, the CNRS-IN2P3 and the Astroparticle Interdisciplinary Programme of the CNRS, the U.K. Science and Technology Facilities Council (STFC), the IPNP of the Charles University, the Polish Ministry of Science and Higher Education, the South African Department of Science and Technology and National Research Foundation, and by the University of Namibia. We appreciate the excellent work of the technical support staff in Berlin, Durham, Hamburg, Heidelberg, Palaiseau, Paris, Saclay, and in Namibia in the construction and operation of the equipment.











\end{document}